\documentstyle[aps]{revtex}

\twocolumn

\begin{document}
\title{Quantum theory of incompatible observations }
\author{Z. Hradil{\cite{optika}} and J. Summhammer}
\address{Atominstitut der \"{O}sterreichischen Universit\"{a}ten,\\
Stadionallee 2,\\
A--1020 Wien, Austria}
\date{\today}
\maketitle

\begin{abstract}
Maximum likelihood principle is shown to be the best measure for
relating the experimental data  with the predictions of quantum
theory.
\end{abstract}


Quantum theory  describes events on the  most
fundamental level currently available. The synthesis of information
from mutually incompatible quantum measurements
plays the key role in testing the structure of the theory.
The purpose of this Letter is to show a unique relationship between
quantum theory and the mathematical statistics used to obtain optimal
information from incompatible observations:
Quantum theory prefers the relative entropy (maximum likelihood
principle) as the proper measure for  evaluation of the distance
between measured data and probabilities defined by quantum theory.

    In the standard textbooks \cite{Sakurai}, a quantum  measurement
is represented by  a hermitian operator $ \hat A $, whose spectrum
determines the possible results of the measurement
\begin{equation}
\hat A |a\rangle = a |a \rangle.
\end{equation}
In the following, the Dirac notation will be used and for the sake
of simplicity a discrete spectrum will be assumed.
Eigenstates are orthogonal $\langle a | a' \rangle = \delta_{aa'}$
and the corresponding projectors provide the closure relation
\begin{equation}
\sum_a |a\rangle\langle a |  = \hat 1.
\end{equation}
Projectors predict the probability  for detecting a particular
value of the q--variable  $a$ represented by the operator $\hat A$
as  $ p_a = \langle a| \rho |a\rangle , $ provided that the system
has been prepared in a quantum state $\rho.$ This mathematical
picture corresponds to the  experimental reality in the following
sense:  When the measurement represented by the operator $\hat A$
is repeated N times on identical copies of the system, the number a
particular output $a$  is collected  $N_a$ times. The relative
frequencies
$
f_a = \frac{N_a}{N}
$
 will  sample the true  probability  as $f_a  \rightarrow p_a$
 fluctuating  around them. The  exact values are reproduced only in
the asymptotical limit $N\rightarrow \infty.$ Experimentalist's
knowledge may be expressed in the form of a diagonal density matrix
\begin{equation}
\hat \rho_{est} = \sum_a f_a  |a\rangle \langle a|,
\label{matice}
\end{equation}
 provided that error bars of the order $1/N$ are associated with
the sampled relative frequencies. This should be understood as mere
rewriting of the experimental data $\{N, N_a\}$. Similar knowledge
may be obtained by observations, which can be parameterized by
operators diagonal in the $ |a\rangle $ basis, i.e. by operators
commuting with operator $ \hat A .$ But the possible measurement of
{\it non--commuting} operators yields new information, which cannot
be derived from the measurement of $\hat A$.

From this  viewpoint  it seems to be advantageous to consider the
sequential synthesis of various non--commuting observables. In this
case several operators $\hat A_j,$ $j=1,2,\ldots$  will be measured
by probing of the system $N$ times  together. Now, one expects to
gain more than just the knowledge of  the diagonal elements of the
density matrix  in some a priori  {\em given}  basis. This
sequential measurement of non--commuting observables  should be
distinguished from the similar problem of
 approximate simultaneous measurement of non--commuting
 observables \cite{noncommuting}.
As in the case of the
 measurement of a hermitian operators, the result of sequential
measurements of non--commuting operators may be represented  by  a
series of projectors $|y_i\rangle \langle y_i |. $ This  should  be
accompanied by relative frequencies $f_i$  indicating how many
times a particular output $i $ has been registered, $\sum_i f_i =
1.$  Various  states need not be orthogonal $\langle y_i |y_j
\rangle \neq \delta_{ij}$, in contrast to the previous case of a
hermitian operator. However, this substantial difference has its
deep consequences. The result of the measurement cannot be
meaningfully represented in the same manner as previously. For
example, direct  linking of probabilities  with relative
frequencies  used in standard reconstructions \cite{rec,Welsch}
$
 \rho_{ii} = f_i,
$
$\rho_{ii} = \langle y_i | \hat \rho | y_i\rangle,$
may appear as inconsistent, since
the system of linear  equation is
overdetermined, in general.

A novel approach will be  suggested here. Let us assume the
existence of a quantum measure $F(\rho_{ii}|f_i)$, which
parameterizes the distance  between measured data and probabilities
predicted by  quantum theory. We will search  for the state(s)
located in the closest neighborhood of the data. A general state
may be parameterized in its diagonal basis as
\begin{equation}
\hat \rho = \sum_i r_i |\varphi_i\rangle \langle \varphi_i|.
\label{density}
\end{equation}
The  equation for the extremal states
 may be found analogously to
the treatment developed in \cite{hradil} for maximum likelihood estimation as
\begin{equation}
\sum_i \frac{\partial F}{\partial \rho_{ii}} |y_i\rangle \langle
y_i |  \hat \rho = \lambda  \hat \rho,
\label{eq}
\end{equation}
where $\lambda $ is a Lagrange multiplier. The normalization
condition  ${\rm Tr} \hat \rho = 1 $  sets its value to
$$ \lambda =   \sum_i \frac{\partial F}{\partial \rho_{ii}}  \rho_{ii}. $$
Any composed function $G(F(\rho_{ii}|f_i)) $  fulfills the
same extremal equation (\ref{eq}) with the Lagrange multiplier
rescaled  as  $ \lambda \frac{d G}{d F}.$
Without loss of generality it is therefore  enough to consider the
normalization condition
 $\lambda = 1.$

The extremal equation (\ref{eq})
has the form of a  decomposition of the identity operator on the
subspace, where the density matrix is defined by
\begin{equation}
\sum_i \frac{\partial F}{\partial \rho_{ii}} |y_i\rangle \langle
y_i |  =  \hat 1_{\rho}.
\label{identity}
\end{equation}
This resembles the definition of a probability-valued operator
measure (POVM) characterizing a generalized measurement
\cite{Helstrom}. To link the  above extremalization with quantum
theory, let us postulate the natural  condition for the quantum
expectation value
\begin{equation}
Tr\biggl(  \frac{\partial F}{\partial \rho_{ii}} |y_i\rangle \langle
y_i |  \hat \rho \biggr)  = f_i.
\label{POVM}
\end{equation}
This assumption seems to be reasonable. The synthesis of
sequential non--compatible observations may be regarded  as
a new measurement scheme,
namely the measurement of the quantum state.

The quantum measure  $F$ then fulfills the differential equation
\begin{equation}
  \frac{\partial F}{\partial \rho_{ii}} \rho_{ii}  =
 f_i.
\label{rovnice}
\end{equation}
and  singles out the  solution
in the form
\begin{equation}
F(\rho_{ii}|f_i) =  \sum_i f_i \ln \rho_{ii}.
\label{loglik}
\end{equation}
This is  nothing else than the log likelihood or Kullback--Leibler
relative information  \cite{KL}.
 Formal requirements of quantum theory, namely the interpretation
of the extremal equation as a POVM, result in the concept of
maximum likelihood  in mathematical statistics. The analogy between
the standard  quantum measurement associated with a single
hermitian operator, and a series of sequential measurements
associated with many non--commuting operators is apparent now.  The
former determines the diagonal elements in the basis of orthonormal
eigenvectors, whereas the latter estimates not only the diagonal
elements, but the diagonalizing basis itself. This is the
difference  between measurement of  quantum observable $\hat A $
and  measurement of the quantum state. In this sense maximum
likelihood  estimation may be considered as a new quantum
measurement. The observed quantum state  is given by the solution
of the nonlinear operator equation
\begin{equation}
\sum_i \frac{f_i}{\rho_{ii}} |y_i\rangle \langle
y_i |  \hat \rho =  \hat \rho,
\label{equation}
\end{equation}
which is, in fact, the completeness relation of a POVM
with measured outputs  $\{ f_i\}.$
Special cases  of the solution  (\ref{equation})
 have been discussed recently  for the phase
\cite{phase}, the diagonal elements of the density matrix
\cite{diagonal} and the reconstruction of the  $1/2$ spin  state
\cite{spin}.

Quantum interpretation offers a new viewpoint on the maximum
likelihood estimation. This method is  customarily considered as
just one of many estimation methods, unfortunately one of the most
complicated ones. It is often considered as a subjective method,
since likelihood quantifies  the degree of belief  in a certain
hypothesis. Any physicist, an experimentalist above all,
 would perhaps use as the first choice the least--squares method
 for fitting   theory and data \cite{Welsch}.
 Let us evaluate this as an illustrative counterexample.  In this case
 $F(\rho_{ii}|f_i) = \sum_i (\rho_{ii}- f_i)^2$ and the extremal
 equation reads
 \begin{eqnarray}
2 \sum_i (\rho_{ii} - f_i) |y_i\rangle \langle y_i| \hat \rho  =
\lambda \hat \rho,
 \label{LSM}\\
\lambda = 2 \sum_i (\rho_{ii} - f_i)  \rho_{ii}.
\nonumber
\end{eqnarray}
 Equation  (\ref{LSM}) may be again interpreted as a completeness
relation for the POVM
$$\hat E_i = 2 \frac{ (\rho_{ii} - f_i)}{\lambda}
|y_i\rangle \langle y_i|.  $$
 However the expectation value  is a rather complicated
implicit function of the measured data, since
 \begin{equation}
{\rm Tr}(\hat \rho \hat E_i ) = 2 \frac{ (\rho_{ii} -
f_i)\rho_{ii}}{\lambda}.
 \end{equation}
It does not mean that the least--squares method  is incorrect, it
only means that such fitting does not reveal the structure of quantum
measurement. In this sense the maximum likelihood seems to be
unique and exceptional.

There are several fundamental consequences of this result.
According to Fisher's theorem \cite{Fisher}, maximum likelihood estimation
is unbiased and achieves the Cram\'{e}r--Rao bound asymptotically
for large $N\rightarrow \infty.$
As demonstrated here, for any finite  $N$  maximum likelihood may
be interpreted as a quantum measurement. When seen this way,
bias and the noise above the Cram\'{e}r--Rao bound seem to be
unpleasant but natural properties of quantum systems.
Maximum likelihood may set new bounds on distinguishability
related currently  to the Fisher information \cite{separat}.
 Fisher information  corresponds to the Riemannian
distinguishability metrics and  may be  naturally interpreted as
the distance in the Hilbert space. Besides this, fundamental
equations of quantum  theory such as Schr\"{o}dinger,
Klein--Gordon,
  Pauli etc.
and other physical laws  may be derived from the principle of minimum
Fisher information \cite{f,ff}. Since Fisher information only
approximates the behaviour of likelihood in asymptotical limit,
all these features seem to be involved also in the maximum
likelihood principle. However, the latter one is obviously
 stronger since, as
shown here, only the maximum likelihood
is able to reproduce the structure of quantum measurement for finite observations.

Notice also that  maximum likelihood generalizes the notion of POVM
in the following sense. Actual measurements may be (and usually
always are)  incomplete. However, the synthesis of any incomplete
measurements, namely of the original observations represented here
by $ |y_i\rangle \langle y_i|, $  is complete in the subspace,
where the  resolution of the identity  (\ref{identity}) holds. POVM
and estimated quantum state are mutually connected  in  dependence
on
 the type of measurement  and on  its results. In particular, it is
not necessary to  consider only the special scheme for quantum
state observation as for example  the mutually
 complementary eigenbases \cite{WootFields}.

We conclude with a remark that may shed light on why maximum likelihood is peculiar:
Maximum likelihood is perhaps singled out by Nature, because the
non-symmetric fluctuations of data of multinomial distributions, which are
the results of quantum measurements, are compensated, so to say, by an
equally non-symmetric attribution of degrees of belief to the various
test states. Maximum likelihood takes into account
that, in {\it finite} observations, improbable events tend to appear more
frequently than they should, and conversely, very probable events
tend to appear somewhat less often.

 This work was supported by TMR Network ERB FMRXCT 96-0057
``Perfect Crystal Neutron Optics'' of the European Union and by the
grants of Czech Ministry of Education VS 96028 and CEZ J14/98.


\begin{references}
\bibitem[*]{optika}  Permanent address: Department of Optics, Palack\'{y}
University, 17. listopadu 50, 772 07 Olomouc, Czech Republic.


\bibitem{Sakurai}
J. J. Sakurai, {\it Modern Quantum Mechanics},
Addison--Wesley, 1994.


\bibitem{noncommuting}
The approximate simultaneous measurement of noncommuting
operators $[\hat A,\hat B] \neq 0$
 can always be represented by measurement of commuting operators
$\hat {\cal A} $,$\hat {\cal B}$ defined  on  the extended Hilbert
space
 ${\cal H} = H_s \otimes H_a, $
where $H_s, H_a$ is  the space of  original system and space of
 auxiliary field (ancilla), respectively.



\bibitem{rec}  U. Leonhardt, {\em Measuring of the Quantum State of Light},
Cambridge Press, 1997.

\bibitem{Welsch}
 D.-- G. Welsch, W. Vogel, T. Opatrn\'{y}, ``Homodyne
Detection and Quantum State Reconstruction'',  in {\em Progress in
Optics, 39 }, ed. E. Wolf.



\bibitem{hradil}  Z. Hradil, Phys. Rev. {\bf A 55}, R1561 (1997);
 Z. Hradil, J. Summhammer, H. Rauch, Phys. Lett. {\bf A 261} (1999)
 20.


\bibitem{Helstrom}  C. W. Helstrom, {\em Quantum Detection and Estimation
Theory,} Academic Press, New York 1976.



\bibitem{KL}
  S. Kullback, R. A. Leibler, Ann. of Math. Stat. {\bf 22}, 79
 (1951). Notice the asymmetry between the arguments $f$ and $p$ in
definition of Kullback--Leibler relative information   $ K(f/p)
= \sum_i f_i \ln(f_i/p_i). $  In the paper
 of B. R. Frieden, in {\em Maximum Entropy and Bayesian Methods in
 Inverse Problems}, edited by C. R. Smith, W. T. Grandy Jr.
 (Reidel, Dordrecht 1985), p. 133 is the term Kullback--Leibler
 norm used for opposite ordering  of data and probabilities. The
 case discussed here  is called  generalized Burg principle.



\bibitem{phase}
J. \v{R}eh\'{a}\v{c}ek, Z. Hradil, M. Zawisky, S. Pascazio, H.
Rauch, J. Pe\v{r}ina, Phys. Rev. {\bf A  60}, 473  1999.


\bibitem{diagonal}
  Z. Hradil, R. My\v{s}ka, acta phys. slov. {\bf
48}, 199 (1998).


\bibitem{spin}
Z. Hradil, J. Summhammer, G. Badurek, H. Rauch, Spin state
reconstruction, unpublished.


\bibitem{Fisher}
R. A. Fisher, Proc. Camb. Soc. {\bf 22}, 700 (1925).

\bibitem{separat}
W. K. Wootters, Phys. Rev. {\bf D 23}, 357 (1981);
S. L. Braunstein, C. M. Caves,  Phys. Rev. Lett.  {\bf  72}, 3439
(1994).

\bibitem{f}
M. Reginatto, Phys. Rev. {\bf A 58}, 1775 (1998);
M. Reginatto, Phys. Lett. {\bf A 249}, 355 (1998).

\bibitem{ff}
B. Roy Frieden, {\em Physics  from Fisher Information}, Cambridge Univ. Press,
Cambridge 1999.


\bibitem{WootFields}
W. K. Wootters, Found. Phys. {\bf 16}, 391 (1986).



\end{references}
\end{document}